\documentclass[12pt,preprint]{aastex}
\pdfoutput=1
\shorttitle{TIME DEPENDENT IONIZATION IN MAGNETIC RECONNECTION}
\shortauthors{Imada et al.}

\begin{document}

\title{ MAGNETIC RECONNECTION IN NON-EQUILIBRIUM IONIZATION PLASMA}

\author{S. \textsc{Imada},\altaffilmark{1} 
I. \textsc{Murakami},\altaffilmark{2} 
T. \textsc{Watanabe} \altaffilmark{3}
H. \textsc{Hara} \altaffilmark{3}
T. \textsc{Shimizu} \altaffilmark{1}
}
\altaffiltext{1}{ Institute of Space and Astronautical Science, Japan Aerospace Exploration Agency, 3--1--1 Yoshinodai, Chuo-ku, Sagamihara-shi, Kanagawa 252--5210, Japan}
\altaffiltext{2}{ National Institute for Fusion Science,
  322--6 Oroshi-cho, Toki, Gifu 509--5292, Japan}
\altaffiltext{3}{ National Astronomical Observatory of Japan,
  2--21--1 Osawa, Mitaka-shi, Tokyo 181--8588, Japan}

\begin{abstract}
  We have studied the effect of time-dependent ionization and recombination processes on magnetic reconnection in the solar corona. Petschek-type steady reconnection, in which model the magnetic energy is mainly converted at the slow-mode shocks, was assumed. We carried out the time-dependent ionization calculation in the magnetic reconnection structure. We only calculated the transient ionization of iron; the other species were assumed to be in ionization equilibrium. The intensity of line emissions at specific wavelengths were also calculated for comparison with {\it Hinode} or other observations in future. What we found is as follows: (1) iron is mostly in non-equilibrium ionization in the reconnection region, (2) the intensity of line emission estimated by the time-dependent ionization calculation is significantly different from that with the ionization equilibrium assumption, (3) the effect of time-dependent ionization is sensitive to the electron density in the case that the electron density is less than $10^{10}$ cm$^{-3}$, (4) the effect of thermal conduction lessens the time-dependent ionization effect, (5) the effect of radiative cooling is negligibly small even if we take into account time-dependent ionization.
 
\end{abstract}

\keywords{MHD --- plasmas --- shock waves --- Sun: corona --- Sun: flare --- Sun: UV Radiation}

\section{INTRODUCTION}
Magnetic reconnection has been discussed as one of the important mechanisms for heating and bulk acceleration in astrophysical plasma, because the magnetic field energy can be rapidly released to the plasma during reconnection. One of major aspects of magnetic reconnection is the rapid energy conversion of stored free magnetic energy to kinetic energy, thermal energy, non-thermal particle energy, and wave/turbulence energy. This energy conversion is fundamental and essential to understand the dynamical behavior of plasma \citep[e.g.,][]{zwe} not only in the solar atmosphere (e.g., \cite{pne}) but also in the Earth's magnetosphere \citep[e.g.,][]{hon,nag1,nag2,bau,oie,ima,ima4}, laboratory  \citep[e.g.,][]{baum,ono,yam, ji} , or other astronomical objects. One of the goals for studying magnetic reconnection is to understand how much energies are converted toward plasma and what is happened afterwards. To answer the question, it is essential to observe the entire energy conversion in magnetic reconnection on a large scale continuously. The solar atmosphere is an excellent space laboratory for magnetic reconnection because of its observability of magnetic reconnection on a large scale. 

One of the most famous phenomena associated with magnetic reconnection is the solar flare. Modern telescope observations have confirmed many typical features expected from the magnetic reconnection model. These include cusp-like structure in X-ray images (e.g., \cite{tsu}), non-thermal electron acceleration (e.g., \cite{mas}), chromospheric evaporation (e.g., \cite{ter}), reconnection inflow and outflows (e.g., \cite{yok,inn}), and plasmoid ejection (e.g., \cite{ohy}).  
Recently the {\it Hinode} spacecraft was launched \citep{kos}, and after first light  {\it Hinode} has been revealing many new solar flare aspects. 
The recent observation of magnetic reconnection in solar corona can be summarized as follows. The stored magnetic field energy in the corona before magnetic reconnection \citep[e.g.,][]{kub,mag}, energy release rate \citep[e.g.,][]{jin}, and most forms of energy after magnetic reconnection \citep[e.g.,][and their referce]{ima2,ima4,asa,min} can be estimated in detail. On the other hand, there is not enough observational knowledge of the physical parameters in the reconnection region itself. The inflow into the reconnection region, the temperature of the plasma in the reconnection region, and the fast Alfvenic flows predicted by reconnection, have not been quantitatively measured in sufficient.  {\it Hinode} and/or the {\it Solar Dynamics Observatory} (SDO) may provide some answers if solar cycle 24 ever produces a solar maximum. However, it is important to discuss why most observations cannot detect the predicted flow or temperature in the reconnection region. One of the reasons why we cannot observe inside the magnetic reconnection region is its darkness. Generally we can see the bright cusp-like structure during the solar flare, although the reconnection region, which might be located above the cusp-like structure, is faint. Recently, \cite{ima6} pointed out that ionization cannot reach equilibrium in the magnetic reconnection region because of its fast flow and rapid heating. Actually, the timescale for ionization ($\sim100$ s) is comparable to the Alfven timescale ($\sim100$ s) in magnetic reconnection region. The reconnection region might be much fainter than we expected in some cases. Therefore, it is important to take into account time-dependent ionization process when we interpret the observation of magnetic reconnection region.  

So far, most of the solar observations are discussed with ionization equilibrium assumption. Non-equilibrium ionization was mainly discussed in the category of coronal heating or solar wind formation. \cite{dup} discussed the general characteristics of the ionization balance in the solar transition region and corona when mass outflow is present. They found that the large temperature gradient within the flow can result in a departure from ionization equilibrium. Recently, \cite{ima7} also discuss the time-dependent ionization in the dimming region where the large mass flows were observed. They claimed that ionization equilibrium assumption in the dimming region is violated especially in the higher temperature rage ($\sim 2$MK). \cite{mar} examined the hydrodynamic numerical modeling of ionization state in nanoflare-heated loops and concluded that non-equilibrium ionization can significantly alter the relative ionic abundances in the quiet Sun. Hydrodynamic modeling of the ionization states in nanoflare-heated loops have been studied intensively during several decades, and most results indicates the importance of non-equilibrium ionization in the context not only of comparison between observations \citep[e.g.,][]{rea} but also plasma dynamics itself \citep[e.g.,][]{bra}. Recently, modeling of time-dependent ionization in a post-coronal mass ejection current sheet was also studied by \cite{ko} and \cite{mur}, and they discussed the consistency between the modeling results and the observation. As for the observation, non-equilibrium ionization was studied by using line spectroscopic observation. \cite{kat} studied the time evolution of spectra of He-like \ion{Ca}{19} and \ion{Fe}{24} observed by {\it Yohkoh}/BCS for a solar flare, and found that  the plasma is considered to be ionizing plasma even in the decay phase of the flare. \cite{ima5} discussed the ion thermal temperature (not apparent ion temperature) in an active region from two emission lines of different atomic species (\ion{Fe}{16} and \ion{S}{13}) observed by the EUV Imaging Spectrometer (EIS) onboard {\it Hinode}, and they found that the electron temperature estimated from ionization equilibrium assumption is different from ion thermal temperature in some parts. They claimed that the result may indicate the presence of ionizing plasma.

In this paper, we focus on the effect of time-dependent ionization processes on magnetic reconnection. We have treated numerically the ionization and recombination process in Petschek-type steady magnetic reconnection \citep{pet}. This paper is organized as follows. In the next section, the models and assumptions which we used in our calculation are given. Section 3 is devoted to the results of the time-dependent ionization and its radiation under four of magnetic reconnection conditions. Summary and discussion are given in \S 4.

\section{MODELING}
\subsection{Petschek Reconnection Model}
We have studied the effect of time-dependent ionization processes on magnetic reconnection. Petschek-type steady reconnection was assumed in our study (Figure 1). In this model, the magnetic energy is mainly converted at the slow-mode shocks which extend from the X-line. We defined the size of calculation box is 200$\times$20$\times$200 Mm$^3$, and the outside of the box was assumed to be vacant. The magnetic reconnection X-line is located at (x,y)=(0,0), and the slow-mode shocks are extended from the X-line. We assumed that the upstream and downstream of the slow-mode shocks are uniform in temperature and density. We also assumed all ions and electron have the same flow speed and temperature at the same location. Further, the ion on the not-reconnected magnetic field is assumed to be in ionization equilibrium. The reconnection plane is in the x-y plane, and the depth of reconnection is assumed to be 200 Mm. The reconnection structure is uniform in z direction. We calculated the shock jump condition using the following standard one-dimensinal steady MHD conservation laws in the deHoffmann-Teller frame, in which the electric fields vanishes outside the shock \citep[e.g.,][]{hau};
\begin{equation}
\left[ \rho v_n \right]=0,
\end{equation}
\begin{equation}
\left[ \rho v_n^2 + p + \frac{B_t^2}{2\mu_0} \right]=0,
\end{equation}
\begin{equation}
\left[ \rho v_n v_t - \frac{B_n B_t}{\mu_0} \right]=0,
\end{equation}
\begin{equation}
\left[ \frac{\gamma}{\gamma-1}\frac{p}{\rho}+\frac{1}{2}\left(v_n^2+v_t^2 \right) \right]=0,
\end{equation}
where $\rho$, $v$, $p$, $B$, $\gamma$ are density, velocity, pressure, magnetic field, specific heat ratio, respectively.
The subscripts t and n denote tangential and normal to the shock, respectively.
The square brackets are the usual notation for the difference between the two sides of the discontinuity.
Once the conditions in the upstream and downstream of the slow-mode shock is determined, its location and reference frame also determined  in the reconnection region by $y=\pm \tan\theta_2 x$, where $\theta_2$ ($=\arctan(B_{t2}/B_{n2})$) is the shock angle in the downstream (see, Figure 1b). 
The subscripts 1 and 2 denote upstream and downstream, respectively.
Afterward, the temperature, density, velocity, and magnetic field are also determined in the entire structure.
We have examined the four case of reconnection conditions to discuss the effect of time-dependent ionization. Table 1 shows the jump conditions of the slow-mode shocks in the reconnection region. The way to solve the jump condition is as follows; 1) assume the upstream density ($N_1$), temperature ($T_1$), shock angle ($\theta_1$), and plasma beta ($\beta_1$), and specific heat ratio ($\gamma$), 2)  define the inflow velocity by the assumption that the outflow velocity is equal to Alfven velocity of upstream. Our assumptions on upstream plasma condition are from past observations \citep[e.g.,][]{tsu2,tsu3}. We will discuss the density dependence with Run1-3 and the thermal conduction effect with Run4. In Run1, we assumed the upstream density, temperature, shock angle, plasma beta, inflow velocity, and specific heat ratio, are $10^9$ cm$^{-3}$, 1.5 MK, 85$^\circ$, 0.02, 137 km sec$^{-1}$, and 5/3, respectively. These are normal values for ambient plasma in the solar corona. The other values in Run1 were derived from Rankine-Hugoniot relations. The electron densities in Run2 and 3 are different from Run1. Most of the other values are the same as Run 1. In Run4 we simulated the isothermal shock condition by setting $\gamma \sim 1$. The plasma beta in Run4 is 40 times larger than that in Run1, because temperature in the upstream region is increased by thermal conduction. Although the other values in Run4 are also different from Run1, these values are normal values observed in the solar flare.

\subsection{Non-Equilibrium of Ionization}
In order to study the effect of transient ionization on magnetic reconnection, we have calculated the time evolution of ion charge states. There are many kinds of atomic species in solar corona, and they mainly radiate line emission in ultra-violet wavelength range by bound-bound process. The most dominant element for radiation is iron at coronal temperatures (a few MK). Thus most of the radiative energy loss is from iron line emission. Further, the recent space telescopes such as {\it Hinode}/EIS or {\it SDO}/Atmospheric Imaging Assembly (AIA) mainly observe the emission lines from iron (e.g., \ion{Fe}{9} 171\AA ~or \ion{Fe}{12} 195\AA). Therefore, we concentrated on the time-dependent ionization of iron in this paper.

 The continuity equations for iron is expressed as follows;
\begin{equation}
\frac{\partial n^{Fe}_i}{\partial t}+\nabla \cdot n^{Fe}_i {\bf v}  = 
n_e\left[n^{Fe}_{i+1} \alpha^{Fe}_{i+1}+ n^{Fe}_{i-1} S^{Fe}_{i-1}-n_i^{Fe}\left(\alpha^{Fe}_{i}+S^{Fe}_{i}\right)\right],
\end{equation}
where $n_i^{Fe}$ is the number density of the $i$th charge state of the iron, $\alpha^{Fe}_i$ represents the collisional and dielectronic recombination coefficients, and $S^{Fe}_i$ represents the collisional ionization coefficients. The ionization and recombination rates were calculated using \cite{arn1}, \cite{arn2}, and \cite{maz}. Here we assumed that all ions and electrons have the same flow speed and temperature in the same upstream location. Note that the ions just across the slow-mode shocks have still the charge state distributed at the coronal temperature, although the temperature is already jump up to that of downstream ($T_2$). The ionization and recombination coefficients ($\alpha$ and $S$) strongly depend on temperature and weakly depend on density. The timescale for ionization and recombination is proportional to $n_e^{-1}$ (see Equation 5). 

 We examined the time-dependent ionization calculation in the magnetic reconnection region with the assumption that plasma does not mix with the plasma coming from the other slow-mode shock crossings, 'stream line model' (see, \cite{ko}). The typical outflow velocity in our calculation is roughly 1500 km s$^{-1}$, and iron thermal velocity at $T=31.3$ MK is roughly 100 km s$^{-1}$. It seems that the plasma mixing for iron along the magnetic field line is small, because the thermal velocity of iron is small compared with outflow velocity. Figure 2 shows the example of our time-dependent ionization calculation (Run1) in the magnetic reconnection region. The plasma conditions are in Table 1. The calculation was carried out in the plasma comoving frame. The horizontal axis shows the time from crossing the slow-mode shock, and the vertical axis shows the ionic fraction of iron. Because we assumed that the ions are initially ionization equilibrium ($T_1 = $1.5 MK),  \ion{Fe}{13} is dominant at $t\sim0$ in Figure 2. After crossing the slow-mode shock, plasma rapidly ionizes by the collisions with the hot electrons ($T_2 = $31.3 MK), and \ion{Fe}{25} dominates after 100 seconds from the slow-mode shock crossing. Roughly speaking, ionization equilibrium is accomplished within 10$^3$ seconds in the case that upstream electron desity is 10$^9$ cm$^{-3}$. In the case that the upstream electron density ($N_1$) is equal to 10$^{10}$(Run2)/ 10$^{8}$(Run3) cm$^{-3}$, the ionization equilibrium time scale is changed to 10$^2$/10$^4$ seconds, respectively (not shown here).

\subsection{Thermal Conduction and Radiative Cooling}
Thermal conduction and radiative cooling govern the evolution of the electron temperature in the downstream of the slow-mode shocks. It is generally believed that the thermal conduction effectively works in reconnection region because of its nonlinearity \citep[e.g.,][]{yok3,yok2}. 
Heat conductivity increases with increasing temperature nonlinearly ($\propto T^{5/2}$).
The thermal conduction is anisotropic, working only along the magnetic field line. 
Therefore, the slow-mode shocks in the reconnection become isothermal shocks owing to the thermal conduction. 
Thermal conduction is a time-dependent process, and we need to solve the time-dependent energy equation. 
Actually, the thermal conduction front is propagating along the magnetic field with finite-time. 
However the heating at the slow-mode shock is very strong, and the electron thermal velocity ($\sim$ 10,000 km s$^{-1}$) is enough faster than the typical velocity of reconnection outflow ($\sim$ 1,000 km s$^{-1}$) that we can neglect the finite-time of propagation of the thermal conduction front \citep[see Figure 2 in][]{yok3}.
Thus, we simply solve the isothermal shock condition by setting $\gamma \sim 1$ instead of solving the time-dependent energy equation directly.
We will discuss the effect of thermal conduction in section 3.2.

The radiative cooling timescale is relatively long compared with the dynamical time scale of a solar flare. Therefore, it is generally believed that radiative cooling does not affect the entire plasma dynamics of the solar flare. The radiative cooling process can contribute the dynamics of post flare loops. On the other hand, it is interesting to estimate to what extent time-dependent ionization can affect the radiative cooling process. Therefore, we assumed that the radiative cooling cannot affect the slow-mode shock condition or plasma dynamics in the downstream, but only that electron temperature is changing by radiative cooling in the plasma comoving frame as following equation; 
\begin{equation}
\frac{\partial T_e}{\partial t} = -\frac{2}{3k_B}n_e\Lambda_{(T_e)},
\end{equation}
where $k_B$ is Boltzmann constant, and $\Lambda_{(T_e)}$ is radiative energy loss function. We neglected the energy exchange between ions and electrons. This process might calm down the cooling of electrons. In that sense, we might overestimate the effect of radiative cooling by factor of $\sim$2 in Equation 6. Note that we assumed all ions and electrons have the same temperature. We distinguish them only in the case of evaluating the radiative cooling effect with Equation 6. We calculate the radiative energy loss function by CHIANTI atomic database 6.0 \citep[e.g.,][]{der} but with ionic fractions of iron calculated by Equation 5. Although $\Lambda_{(T_e)}$ also depends weakly on the electron density, we neglect the density dependence and assumed the electron density is 10$^{10}$ cm$^{-3}$. The radiative cooling includes bound-bound, bound-free, and free-free processes. The dominant radiative process in solar corona is bound-bound emission, which we already mentioned above. Because bound-bound emissions are heavily affected by the ionic fraction, the radiative cooling may be sensitive to the time-dependent ionization process. The ions are most likely in non-equilibrium ionization in the downstream of the slow-mode shocks (Figure 2). We assumed all the elements except iron are in ionization equilibrium, because most part of radiative energy loss is from iron in coronal plasma. We used the usual coronal abundance in \cite{fel} to estimate the line emissions. We will discuss the effect of time-dependent ionization on radiative cooling process in section 3.3.

\subsection{Line Emissions in Ultra-Violet}
It is useful to calculate the intensity of line emission at specific wavelengths for comparison with the recent modern observations. We selected the strong emission lines which are sensitive to the hot component such as solar flare plasma in ultra-violet wavelength. Table 2 shows the lines used in our study.  
We also calculated \ion{Fe}{12} to monitor the plasma in the upstream of the slow-mode shocks.
In a low-density plasma such as the solar corona, the processes that populate and depopulate the excited levels of an ion are generally much faster than the processes that are responsible for ionization and recombination. Thus we assumed that level populations are always in equilibrium in each plasma condition. We calculated each line emissions by CHIANTI atomic database 6.0 with ionic fraction estimated by Equation 5. Further we assumed that the contribution functions do not depend on density, because the dependences are very small in our situations.

\section{Results}
\subsection{Standard Petschek-type Magnetic Reconnection}
Figure 3 shows the intensities of the specific wavelength in the magnetic reconnection region (Run1). The horizontal axis shows x, and the vertical axis shows y in Figure 1. The colors show the intensities of \ion{Fe}{12}, \ion{Fe}{18}, \ion{Fe}{19}, \ion{Fe}{20}, \ion{Fe}{21}, \ion{Fe}{22}, \ion{Fe}{23}, \ion{Fe}{24} (Table 2). We show the results only x$>$0. The magnetic reconnection X-line is located at (x,y)=(0,0), and the pair of slow-mode shocks are extended from the X-line. The results of time-dependent ionization are shown in y$>$0. We also show the results of ionization equilibrium in y$<$0 for comparison. The line-of-sight (LOS) depth is assumed to be 200 Mm, and the intensities linearly depend on LOS depth.
Note that the thermal conduction or radiative cooling effects are not included in this calculation.
 
We can clearly see that there are typically two regions which are bright and dark in the result of \ion{Fe}{12}. The boundaries for bright and dark regions correspond to the location of the slow-mode shocks. The ionization timescale for \ion{Fe}{12} at $T_e$=31.3 MK is very short (see Figure 2). Therefore, the difference between equilibrium (y$<$0) and non-equilibrium (y$>$0) is not clear. The bright broad linear structures can be seen from \ion{Fe}{18} to \ion{Fe}{22} only in non-equilibrium results. These linear structures are parallel to the slow-mode shocks. The ionization timescales for those ions are longer than that for \ion{Fe}{12}. Therefore, the ions, which are still ionizing, are advected by the fast outflow ($\sim$1500 km s$^{-1}$) toward the deep downstream of the slow-mode shocks. For example, in Figure 2 the \ion{Fe}{19} population can achieve a peak in a ten seconds (15 Mm) and reduce down to less than 0.1\% within a hundred seconds (100 Mm). On the other hand, we cannot see any line emissions in the case of ionization equilibrium results. The electrons are heated by crossing the slow-mode shocks, and at the same time and place the ions are ionized and achieve ionization equilibrium in y$<$0 of Figure 3. In the ionization equilibrium condition, the population of  \ion{Fe}{18} to \ion{Fe}{22} is less than 1\%. Thus the emissions from those ions are negligibly small. For \ion{Fe}{23} and \ion{Fe}{24} emissions in non-equilibrium conditions are stronger than that in equilibrium conditions. The peak ionic population for those ions are achieved in $\sim$50 seconds. The flow transit timescale of  our calculation box is also $\sim$50 seconds. Therefore, the ionic fraction of \ion{Fe}{23} and \ion{Fe}{24} are almost at the peak, and the emissions are very strong. 

The ionization timescale is proportional to the electron density. Therefore the high-density condition causes the lessening of the time-dependent ionization effect. Figure 4 shows the emissions under relatively high-density conditions (Run2). The figure format is the same as Figure 3. Note that the color scale is 100 times higher than that in Figure 2, because the intensity itself is proportional to $n_e^2$. The bright linear structures in Figure 4, from \ion{Fe}{18} to \ion{Fe}{22}, become narrower than that in the case of $N_1\sim 10^9$ cm$^{-3}$. The ionization timescale in the high-density plasma ($N_1$=10$^{10}$ cm$^{-3}$) is one-tenth of that in Run1, although the dynamical timescale does not change. Therefore, the ions can achieve ionization equilibrium on a shorter spatial scale than that in Run1. We can clearly see that most of the intensity is the same between in y$>$0 and y$<$0. This indicates that the ionization timescale is very short compared with dynamical timescale in Run2. The difference between y$>$0 and y$<$0 can be seen only around the slow-mode shocks.

 It is plausible that the low-density condition causes the strengthening of the time-dependent ionization effect. Figure 5 shows the result of transient ionization in the low-density plasma ($N_1$=10$^{8}$ cm$^{-3}$, Run3). In the low-density condition, the emissions from \ion{Fe}{21} to \ion{Fe}{24} are quite small in the time-dependent ionization results. The emissions from \ion{Fe}{18} to \ion{Fe}{20} are strong, because the ionic fraction is roughly around the peak in our calculation box. Further we can clearly see the enhancement of \ion{Fe}{12} emissions in the result of time-dependent ionization, which cannot be seen in Run1 and 2. This enhancement is caused by the compression at the slow-mode shock. The density in the downstream of the slow-mode shocks are $\sim$2.5 times as much as that in the upstream. The same enhancement should occur even in Run 1 and 2, though the ionization time-scale is too short to detect the enhancement of \ion{Fe}{12}.

We discussed the time-dependent ionization effect in x-y plane (reconnection plane, Figure 1a) with Figure3-5.  It is useful to change the LOS direction parallel to x axis. We assumed all physical values are uniform in the z direction. Thus we show the spatial variation only  in the y direction in Figure 6. The magnetic reconnection conditions are the same as Run1. The horizontal axis shows y, and the vertical axis shows the intensity of each line emission integrated along the LOS (-100$<$x$<$100). The results of time-dependent ionization are shown in $y>0$, and the result of the ionization equilibrium assumption are demonstrated in $y<0$. The intensity distribution for upstream of the slow-mode shock is represented by \ion{Fe}{12} in Figure 6. The intensity of \ion{Fe}{12} is almost symmetric in the y direction, and their minimum/maximum is located around $|$y$|\sim$0/10, respectively. These are not from time-dependent ionization but from the geometry of the slow-mode shocks. The spatial distribution of \ion{Fe}{12} is not much different between non-equilibrium and equilibrium ionization, because the ionization timescale is very short. On the other hand, from \ion{Fe}{18} to \ion{Fe}{23}, the intensity calculated from our time-dependent ionization is stronger than that calculated with ionization equilibrium assumption for the most part. The intensity of \ion{Fe}{24} is roughly the same in both cases, although the gradient of decreasing from the peak is much steeper in the non-equilibrium ionization results.

It is useful to display the line spectrum radiated from the reconnection region to compare with recent flare observations. Figure 7 shows the line spectrum of \ion{Fe}{24} along y$=$0 in Figure 6. The horizontal axis shows the wavelength, and the vertical axis shows the spectral intensity. The line center position for stationary \ion{Fe}{24} are represented by dashed lines. The line width is determined only by the thermal velocity of the ions. Although in actual observation the instrumental or non-thermal width can contribute to broadening the line spectrum, we neglect them for the simplicity. The two line components of \ion{Fe}{24} can be seen around 191 and 193 \AA, because of the bi-directional fast reconnection outflows (1500 km s$^{-1}$). The line components are completely separated, because the thermal velocity is sufficiently small compared with the reconnection outflow.

\subsection{Thermal Conduction}

In section 3.1 we neglect thermal conduction along the magnetic fields in the reconnection region. It is generally believed that the slow-mode shocks in the reconnection become isothermal shocks owing to the thermal conduction \citep{yok3,yok2}. Thermal conduction is a time-dependent process, and we need to solve the time-dependent energy equation. We simply solve the isothermal shock condition by setting $\gamma \sim 1$ instead of solving the energy equation directly. The shock jump conditions are in Run4 of Table 1. In this calculation, the outflow velocity is relatively slow (780 km s$^{-1}$), because the Alfven velocity in upstream is slower than that in Run1-3.

Figure 8 shows the emissions from the magnetic reconnection region which includes the thermal conduction effect. We also take into account the time-dependent ionization process in this figure. We can also see that there are bright and dark region in \ion{Fe}{12}. The boundary for bright and dark regions corresponds to the magnetic field which connects to the magnetic reconnection X-line (magnetic separatrix). Because the electrons upstream of the slow-mode shock are also heated up to 15 MK, the ionization process can proceed even in the upstream. The slow-mode shock in Figure 8 is located in the same position in Figure 3. We can clearly see the hot plasma between the magnetic separatrix and the slow-mode shock, so called 'thermal halo', from \ion{Fe}{18} to \ion{Fe}{24}.  The intensity downstream of the slow-mode shock is strong and the ionization proceeds much faster owing to the density compression at the slow-mode shock.

It is useful to display the line spectrum from the reconnection with thermal conduction. Figure 9 shows the line spectrum of \ion{Fe}{24} along y$=$8.5 in Figure 8. The figure format is the same as Figure 7. The two line emissions of \ion{Fe}{24} can be seen around 191.5 and 192.5 \AA, because of the bi-directional fast reconnection outflows (780 km s$^{-1}$) in Run4. The emissions from the 'thermal halo' also can contribute to the line profile, and we can clearly see the stationary component in \ion{Fe}{24}. Even in Run4 the line components are completely separated, because the thermal velocity is sufficiently small compared with the reconnection outflow.

\subsection{Radiative Cooling}
To understand to what extent time-dependent ionization can affect the radiative cooling process, we calculate the time evolution of radiative energy loss rate and electron temperature with the time-dependent ionization in the magnetic reconnection region. Figure 10 shows the radiative energy loss rate and electron temperature along y$=$0 in Run1. The horizontal axis shows the distance from the X-line, and the vertical axis shows the radiative energy loss rate and electron temperature. The solid lines show the results with the time-dependent ionization process, and the dashed lines show the results with the ionization equilibrium assumption. In this calculation electron temperature varies with time even after crossing the slow-mode shock. This naturally causes the ionization and recombination coefficients ($\alpha$ and $S$) to vary with time. Therefore, in this calculation we need to solve the time-dependent ionization coupled with radiative cooling.

The radiative energy loss rate estimated by the time-dependent ionization calculation is much larger than that  with the ionization equilibrium assumption everywhere in the reconnection region. In the non-equilibrium ionization case, the radiative energy loss rate is peaked around $x=5$ Mm and the absolute value is 1.4$\times 10^{-22}$ erg sec$^{-1}$ cm$^{3}$. On the other hand, in the ionization equilibrium case, the radiative energy loss rate is almost flat and the absolute value is 2.8$\times 10^{-23}$ erg sec$^{-1}$ cm$^{3}$. Therefore the effect of radiative cooling can be stronger  in the case that time-dependent ionization is taken into account. However, the cooling itself is very weak. The cooling time scale (for example, $\delta t = 3k_BT_e/2n_e\Lambda$ $\sim$ 6 hour at $T_e=30$MK, $n_e=3\times10^9$cm$^{-3}$, $\Lambda=10^{-22}$ erg sec$^{-1}$ cm$^3$) is still much longer than the Alfven timescale (100 sec). Therefore, the cooling of electron temperature is negligibly small even in the case of non-equilibrium conditions. One may think that the radiative cooling effectively work in  the high density condition, such as $N_1 \sim 10^{10} $cm$^{-3}$. However even under such a condition the radiative cooling effect is still weak. We will discuss this point later.  

\section{Summary and Discussion}
We have studied the effect of time-dependent ionization and recombination on magnetic reconnection in the solar corona. We assumed the Petschek-type steady reconnection and calculated the time-dependent ionization in the magnetic reconnection structure. We found that iron is mostly still ionizing downstream of the slow-mode shocks. The intensity of line emissions estimated by the time-dependent ionization calculation is significantly apart from that estimated by the ionization equilibrium assumption. We also found that the effect of time-dependent ionization is sensitive to the electron density in the case that the electron density is less than $10^{10}$ cm$^{-3}$. We also studied the effect of thermal conduction on the ionization process in the reconnection region. We found that thermal conduction caused the lessening of the time-dependent ionization effect, because the ionization process can proceed even upstream of the slow-mode shocks. The faint "thermal halo" also can be observed in the calculation with the thermal conduction, and the observed entire structure is significantly different from that without thermal conduction. The radiative energy loss in the non-equilibrium ionization plasma are also discussed. The effect of radiative cooling is negligibly small even if we take into account time-dependent ionization.

Recently, much work has been done about the atomic database. 
One of the main progress is the updating of recombination rate coefficients for bare through Na-like ions \citep[e.g.,][]{bry}.
The ionization rate coefficients are also updated in part \citep[e.g.,][]{der2}.
The updating of atomic database may affect to some extent on our results.    
Actually, the ionic abundances in the ionization equilibrium might be changed.
However, in our situation, the ionization rate coefficients are quite larger than the recombination rate coefficients.
Thus the effect of the recent updating of atomic database is limited in our case.

Let us discuss the difference between the time-dependent ionization calculation results with and without thermal conduction. 
In Figure 3 (without thermal conduction) we can clearly observe the linear structure which is parallel to the slow-mode shocks. 
Because the density and temperature only can vary when the plasma crosses the slow-mode shocks, the same ionization degree is parallel to the slow-mode shocks. On the other hand, in Figure 8 (with thermal conduction), the ionization can proceed even in the upstream. Thus, the same ionization degree is not parallel to the slow-mode shocks any more. To discuss the difference between the calculation results with and without thermal conduction quantitatively, we define the ionization degree as follows,
\begin{equation}
\tau_{(x,y)} \equiv \sum_i \frac{L_i}{v_i}\frac{n_i}{n_1},
\end{equation}
where $\tau$, $L_i$, $v_i$, and $n_i$ are ionization degree, length along the streaming line, velocity, and density in {\it i}th region, respectively. 
Note that our definition only can apply in the case that the temperature is uniform in every $i$th region.
Only the density can be different in each region. 
The duration of ions staying in $i$th region is represented by ${L_i}/{v_i}$ in Equation 7. 
Because the ionization timescale is proportional to $n^{-1}$, we normalized $i$th density by the upstream density.
The schematic illustration of our variables in Petschek reconnection configuration is at the top of Figure 11.
We also defined the inclination of the magnetic separatrix and the slow-mode shock as $\phi_1$ and $\phi_2$, respectively.
We can derive $L_i$ from the simple geometrical information in Figure 11a.
By using Equation 7, we can simply derive the contour of same ionization degree ($\tau_{(x,y)}$) in the downstream of the slow-mode shock (without thermal conduction) as follows,
\begin{eqnarray}
y=\tan\phi_2 \left(x-v_2\frac{n_1}{n_2}\tau\right) \qquad \quad.
\end{eqnarray}
The slope in Equation 8 is the same as the inclination of the slow-mode shock ($\tan\phi_2$). 
Thus, the same ionization degree is parallel to the slow-mode shocks (Figure 11(b)).
Let us move to the result with the thermal conduction.
In the same way as Equation 8, we can also derive the contour as follows,
\begin{eqnarray}
y=&\left(\frac{1}{\tan\phi_2}-\frac{v_2}{v_1}\frac{n_1}{n_2}\frac{\sin\phi_1}{\sin\phi_2}\right)^{-1} \left(x-v_2\frac{n_1}{n_2}\tau\right)\qquad \quad &{\rm for \quad downstream}
\nonumber\\
=&\tan\left(\phi_1+\phi_2\right)\left(x-\frac{1}{\sin\left(\phi_1+\phi_2 \right)}v_1\tau\right) \qquad \quad &{\rm for \quad thermal \quad halo}.
\end{eqnarray}
We can find that the slope in Equation 9 for the downstream of the slow-mode shock is much steeper than that in Equation 8.
For the 'thermal halo', the slope in Equation 9 is the same as that of the magnetic separatrix ($\tan(\phi_2+\phi_2)$). 
Although the slope for the downstream is slightly steeper than that for the 'thermal halo', the difference between them is negligibly small.
Thus, the same ionization degree is roughly parallel to the magnetic separatrix both downstream and in the 'thermal halo' (Figure 11(c)).
We assumed $\gamma \sim 1$ instead of solving the energy equation directly to include the thermal conduction effect.
The inclination angle of 'thermal halo' might be slightly modified, when we solve the energy equation directly.
Because the thermal conduction front is propagating with finite time, the boundary of 'thermal halo' will be slightly moved from the magnetic separatrix toward the slow mode shock. 

In \S 3.3 we discussed the radiative cooling effect with time-dependent ionization in the magnetic reconnection region. We concluded that the effect is negligibly small in the coronal magnetic reconnection ($n_1=$10$^{9}$ cm$^{-3}$) even if we take account of time-dependent ionization. Let us discuss the radiative cooling effect in high-density plasma.
Generally, the radiative cooling effectively works in dense plasma, because the cooling rate is proportional to the electron density (see Equation 6). Further we found that the radiative loss rate ($\Lambda_{(T_e)}$) considering the time-dependent ionization in reconnection region is significantly larger than that with ionization equilibrium assumption. Thus the electron might be cooling in high-density condition by radiation.  However, the ionization timescale is shorter in dense plasma, which we already mentioned. After all, the enhancement of radiative energy loss in dense plasma is not so large; the radiative cooling with time-dependent ionization is limited. The magnetic reconnection in transition region ($T_e \sim 0.1$MK) might be a different situation. This is for future work.

Anomalies in elemental abundance are often observed during solar flares \citep[e.g.,][]{fel1}. It is useful to estimate how the radiative cooling can be enhanced due to anomalies in elemental abundance. Figure 12 shows the variation of radiative energy loss rate and electron temperature with the assumption that the iron abundance is 100 times higher than the typical observed in the average corona. The figure format is the same as Figure 10.
The radiative loss rate is enhanced 100 times because of the anomalous iron abundance. The electron temperature can be cooled down to $\sim$ 25 MK in the case of non-equilibrium. 
Even in the extreme case, the radiative cooling does not affect much on the dynamics of magnetic reconnection. Therefore, we can conclude that  the radiative cooling cannot affect reconnection dynamics even if we take into account of time-dependent ionization in most coronal condition.

In this paper, we used several assumptions. For example, Petschek-type magnetic reconnection, $T_i=T_e$, and fast thermal conduction.
There are some discussions that Petschek reconnection is unstable unless resistivity increases at the reconnection site \citep[][]{kul, zwe}.
Recently, the effect of $T_i \neq T_e$ condition on the structure of magnetic reconnection is also discussed \citep[][]{lon}. 
Plasma mixing or trapping in non-steady reconnection region, which includes magnetic islands or turbulence, might affect on the time-depend ionization process. Therefore, we think testing time-dependent ionization in non-steady magnetic reconnection is important.
Comparison between the observations of solar flare and the time-dependent ionization results in steady or non-steady magnetic reconnection calculation may reveal which, steady or non-steady, is dominated in solar corona.

\acknowledgments 
The authors thank E. G. Zweibel for fruitful discussions, and the hospitality of University of Wisconsin. This work was partially supported by the Grant-in-Aid for Young Scientists Start-up (21840062), by the Grant-in-Aid for Scientific Research B (23340045), by the JSPS Core-to-Core Program (22001),  by the JSPS fund \#R53 (''Institutional Program for Young Researcher Overseas Visits'', FY2009-2011) allocated to NAOJ, and by the NINS Inter-institute collarborative program for Creation of New
Research Area (Head Investigator: T. Watanabe).

\begin{table*}
\begin{center}
\caption{Slow-Mode Shock Jump Conditions. }
\begin{tabular}{crrrrrrrrrrr}
\tableline\tableline
Run & $N_1$& $T_1$ &$\theta_1$& $\beta_1$ & $V_{in}$ & $\gamma$ & $B_1$& $N_2$ & $T_2$ &$\theta_2$ & $V_{out}$  \\
\tableline
1 &10$^9$  &1.5 &85 &0.02  &137 &5/3&22.8 &2.45$\times$10$^9$ &31.3  & 5.2&1560\\
2 &10$^{10}$ &1.5 &85  &0.02 &137 &5/3&72.1 &2.45$\times$10$^{10}$ &31.3 & 5.2 &1560 \\
3 &10$^8$ &1.5 &85 &0.02 &137 &5/3&7.2& 2.45$\times$10$^8$ &31.3 & 5.2&1560 \\
4 &10$^9$ &15 &85 &0.8 &68 &1.01&11.4 &2.22$\times$10$^9$ &15.2 & 5.2&780 \\
\tableline
\end{tabular}
%% Any table notes must follow the \end{tabular} command.
\tablecomments{$N_1$, $T_1$, $\theta_1$, $\beta_1$, $V_{in}$, $\gamma$, $B_1$, $N_2$, $T_2$, $\theta_2$, and $V_{out}$ are upstream density  (cm$^{-3}$), upstream temperature (MK), upstream shock angle (degree), upstream plasma beta, inflow velocity (km s$^{-1}$), specific heat ratio, upstream magnetic field (G), downstream density (cm$^{-3}$), downstream temperature (MK), downstream shock angle (degree), and outflow velocity (km s$^{-1}$), respectively.}
\end{center}
\end{table*}

\begin{table*}
\begin{center}
\caption{Emission Lines. }
\begin{tabular}{crrrrrrrrrr}
\tableline\tableline
Line & Wavelength (\AA)& logT${\rm _{max}}$(K) & Transition \\
\tableline
\ion{Fe}{12} &195.12  &6.2 &3s2 3p3 4S3/2 - 3s2 3p2 (3P) 3d 4P3/2 \\
\ion{Fe}{18} &974.86 &6.9 & 2s2 2p5 2P3/2 - 2s2 2p5 2P1/2\\
\ion{Fe}{19} &592.24 &7.0 &2s2 2p4 3P2 - 2s2 2p4 1D2 \\
\ion{Fe}{20} &721.56 &7.1 &2s2 2p3 4S3/2 - 2s2 2p3 2D3/2\\
\ion{Fe}{21} &786.16 &7.1 & 2s2 2p2 3P2 - 2s2 2p2 1D2\\
\ion{Fe}{22} &135.79 &7.1 &2s2 2p 2P1/2 - 2s 2p2 2D3/2\\
\ion{Fe}{23} &132.91 &7.2 & 2s2 1S0 - 2s 2p 1P1\\
\ion{Fe}{24} &192.03 &7.2 & 1s2 2s 2S1/2 - 1s2 2p 2P3/2\\
\tableline
\end{tabular}
%% Any table notes must follow the \end{tabular} command.
%%\tablecomments{$a_i$,  standard deviation (\AA).}
\end{center}
\end{table*}

\begin{figure}
\epsscale{0.7}
\plotone{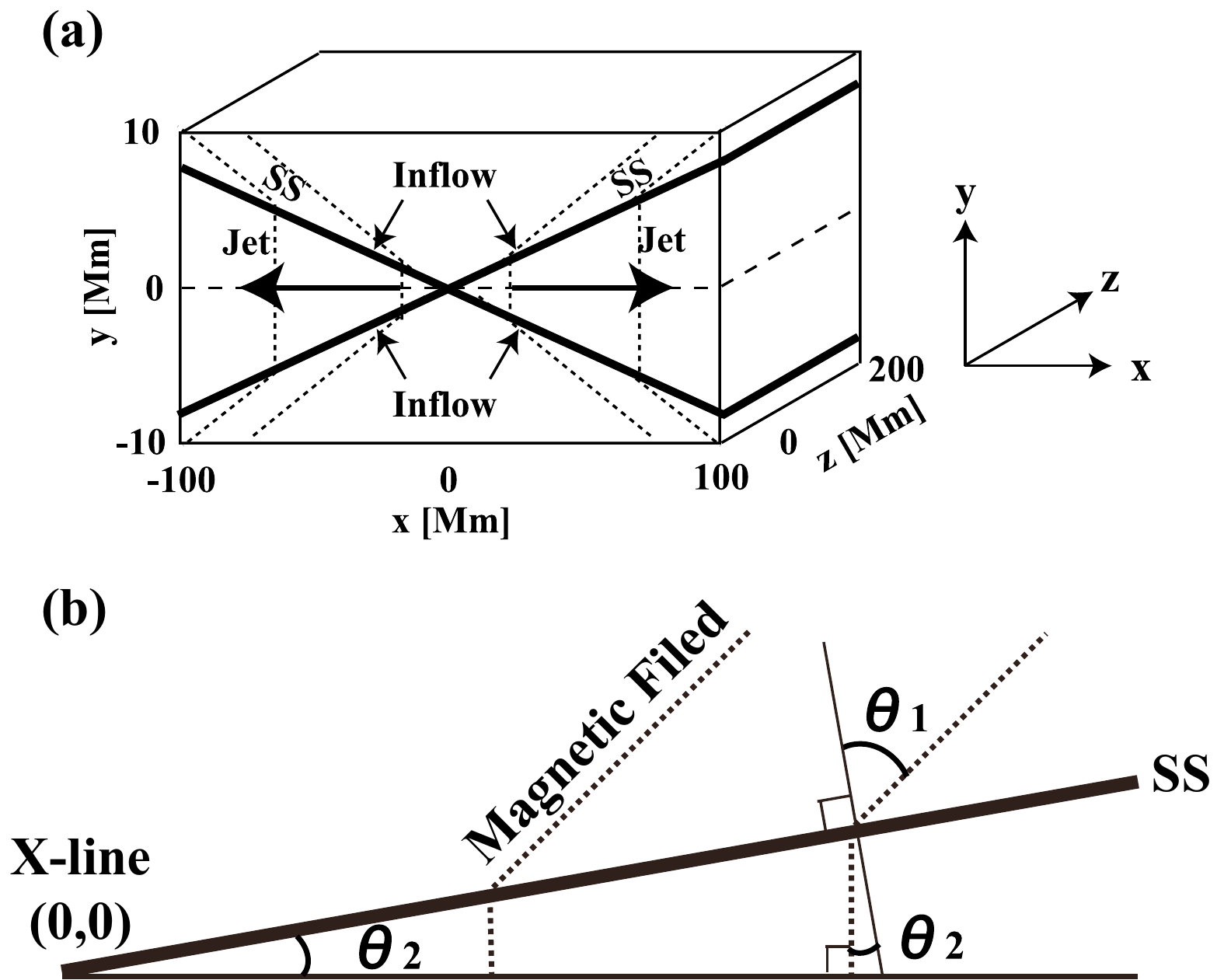}
\caption{ Schematic illustration of Petschek-type steady magnetic reconnection. Dotted lines show magnetic fields and SS are slow-mode shocks. $\theta_1$ and $\theta_2$ show the shock angle of upstream and downstream, respectively. a) entire structure, b)  relationship between shock angle and entire structure. }
\end{figure}

\begin{figure}
\epsscale{0.9}
\plotone{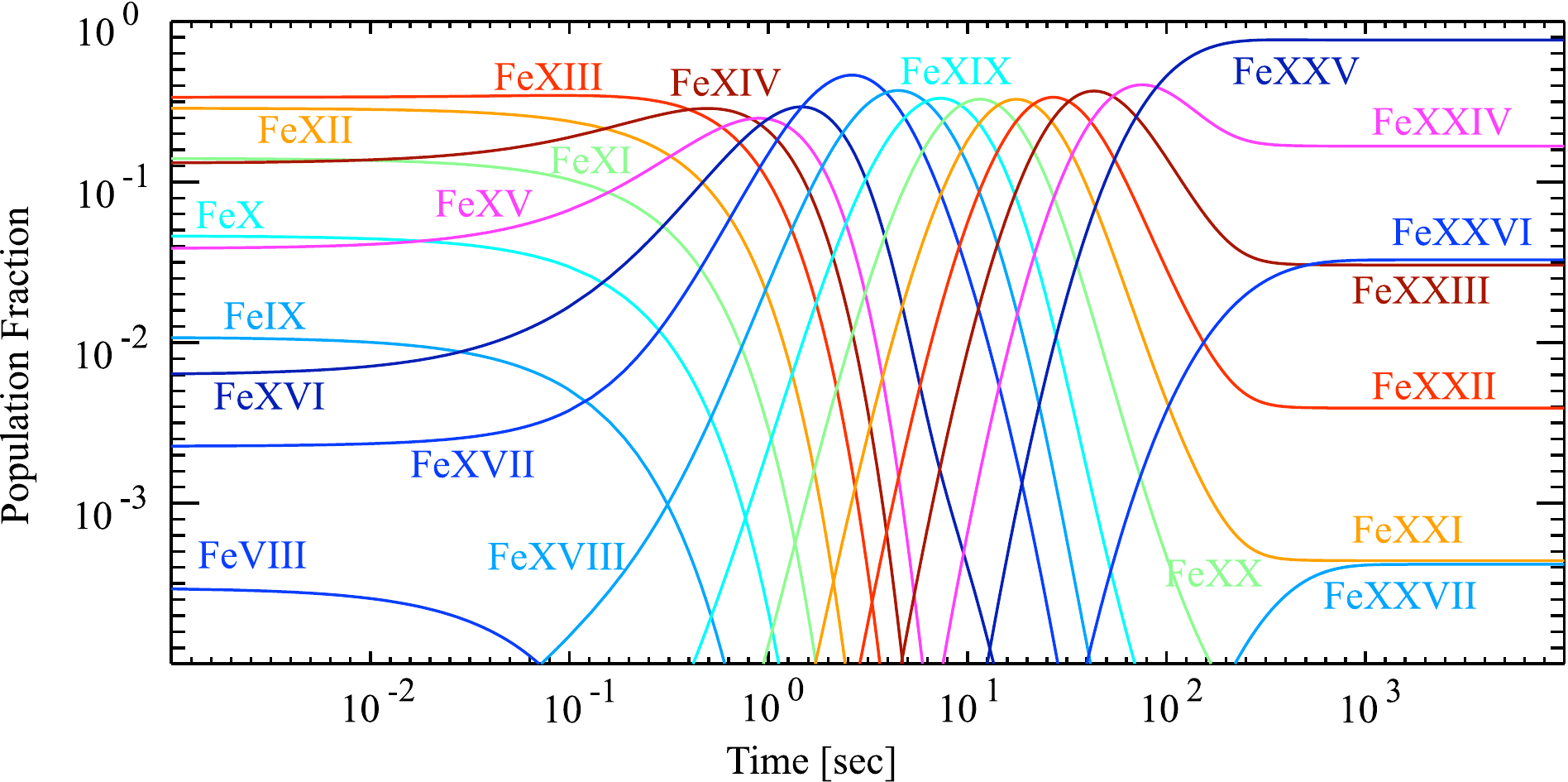}
\caption{ Example of time-dependent ionization in magnetic reconnection (Run1). Time starts from shock crossing. The calculation was carried out in the plasma comoving frame.}
\end{figure}

\begin{figure}
\epsscale{0.9}
\plotone{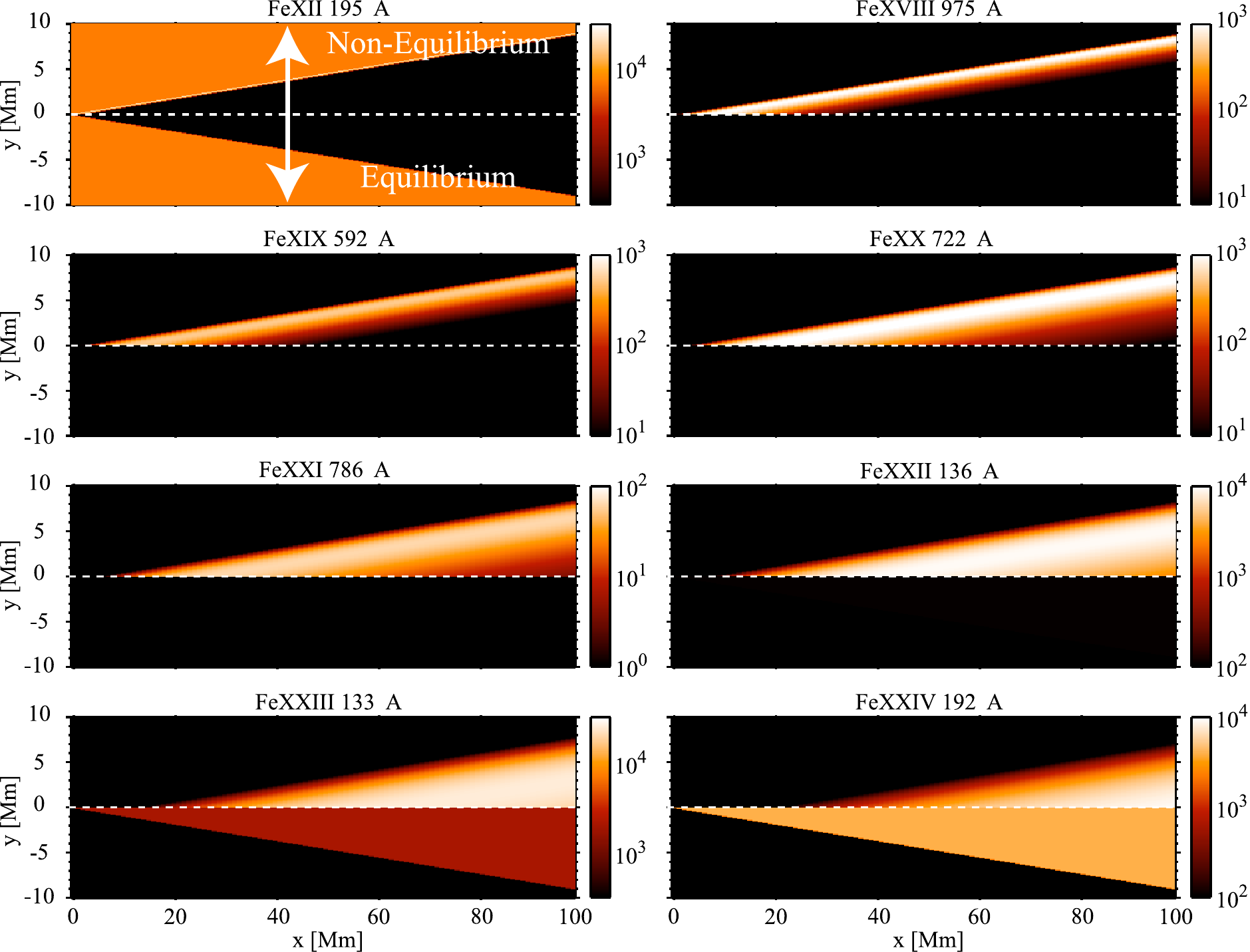}
\caption{ Intensities of \ion{Fe}{12}, \ion{Fe}{18}, \ion{Fe}{19}, \ion{Fe}{20}, \ion{Fe}{21}, \ion{Fe}{22}, \ion{Fe}{23}, and \ion{Fe}{24} from magnetic reconnection region (Run1). The time-dependent ionization results are shown in y$>$0, and the results with ionization equilibrium are shown in y$<0$. Note that the aspect ratio of the figure is different from the real scale.}
\end{figure}

\begin{figure}
\epsscale{0.9}
\plotone{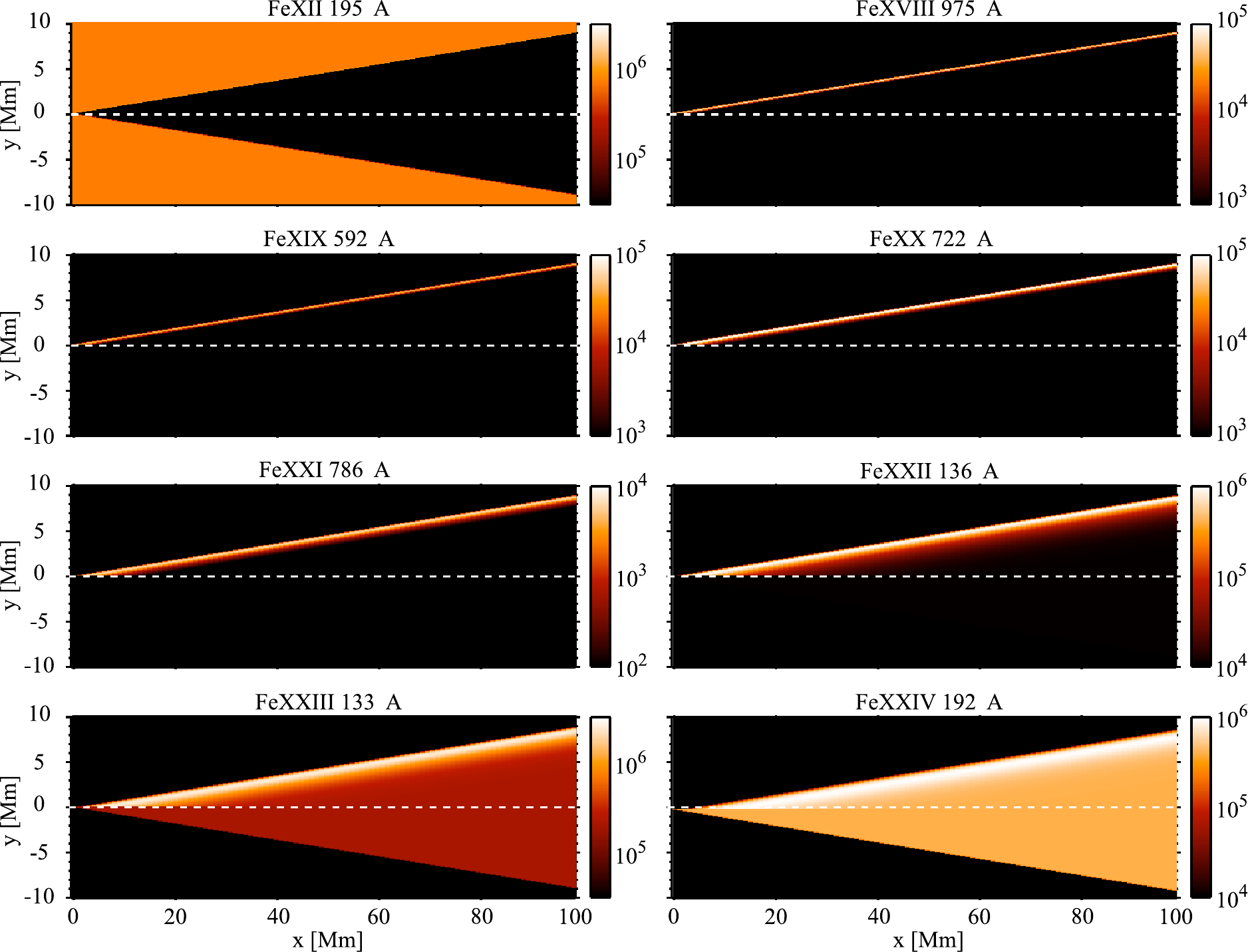}
\caption{ Result of Run2 (high density condition, $N_1$=10$^{10}$ cm$^{-3}$). Figure format is the same as Figure 3. }
\end{figure}

\begin{figure}
\epsscale{0.9}
\plotone{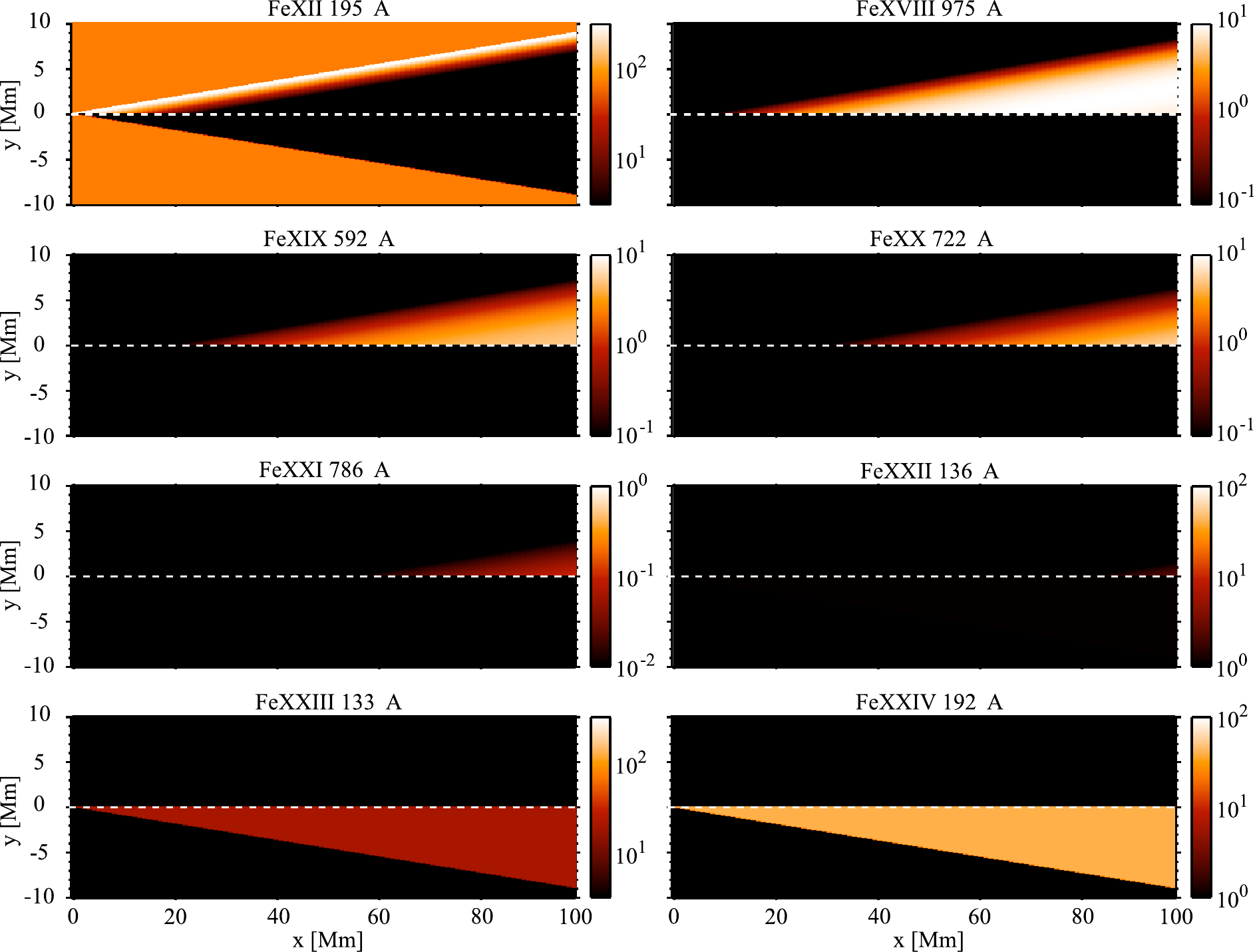}
\caption{ Result of Run3 (low density condition, $N_1$=10$^{8}$ cm$^{-3}$). Figure format is the same as Figure 3.}
\end{figure}

\begin{figure}
\epsscale{0.6}
\plotone{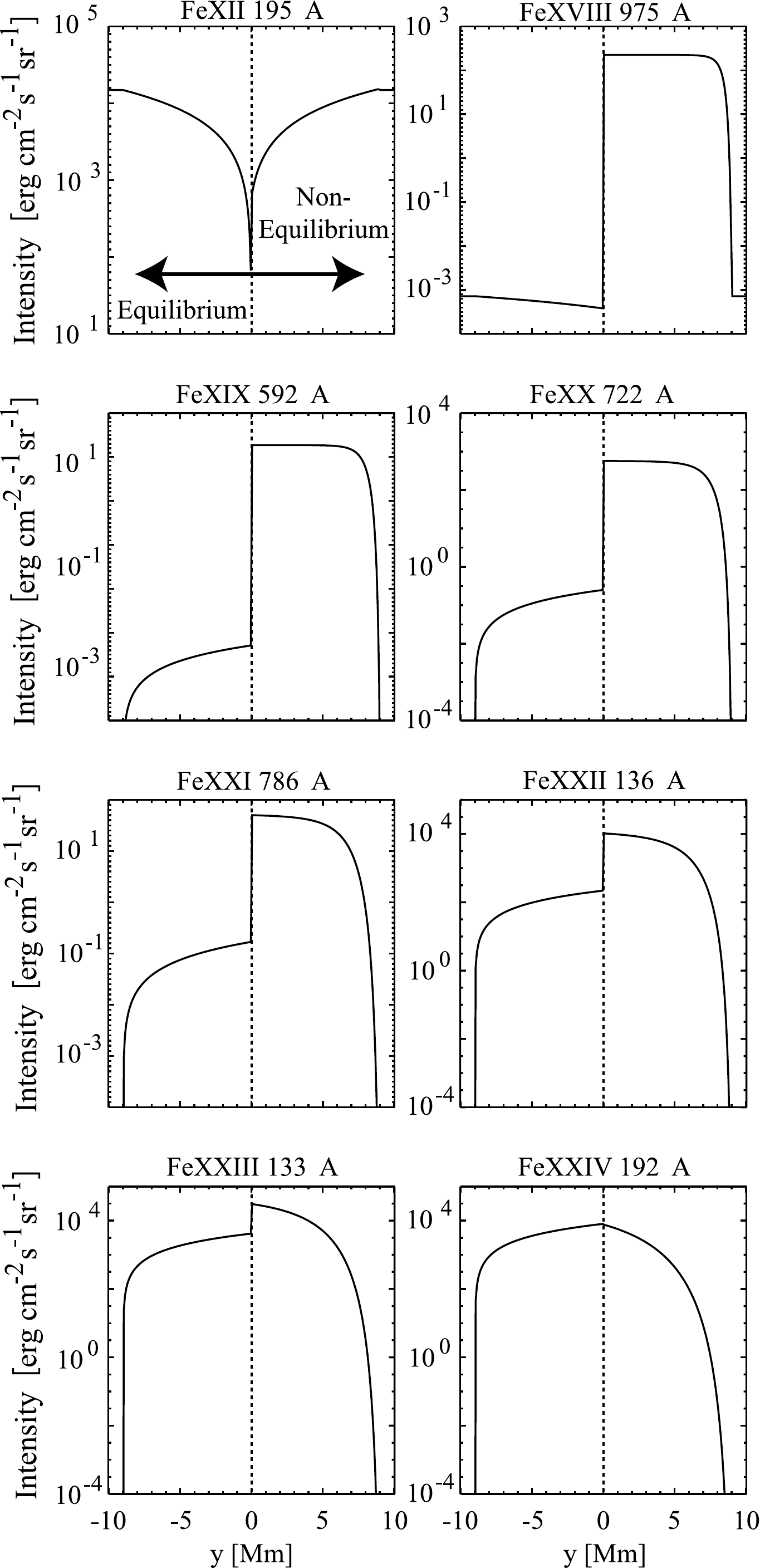}
\caption{ Spatial variations of intensity in the y direction. {\bf The horizontal axis shows y, and the vertical axis shows the intensity of each line emission integrated along the LOS (-100$<$x$<$100). The LOS direction is parallel to the x. The time-dependent ionization results are shown in y$>$0, and the results with ionization equilibrium are shown in y$<0$.} }
\end{figure}

\begin{figure}
\epsscale{0.7}
\plotone{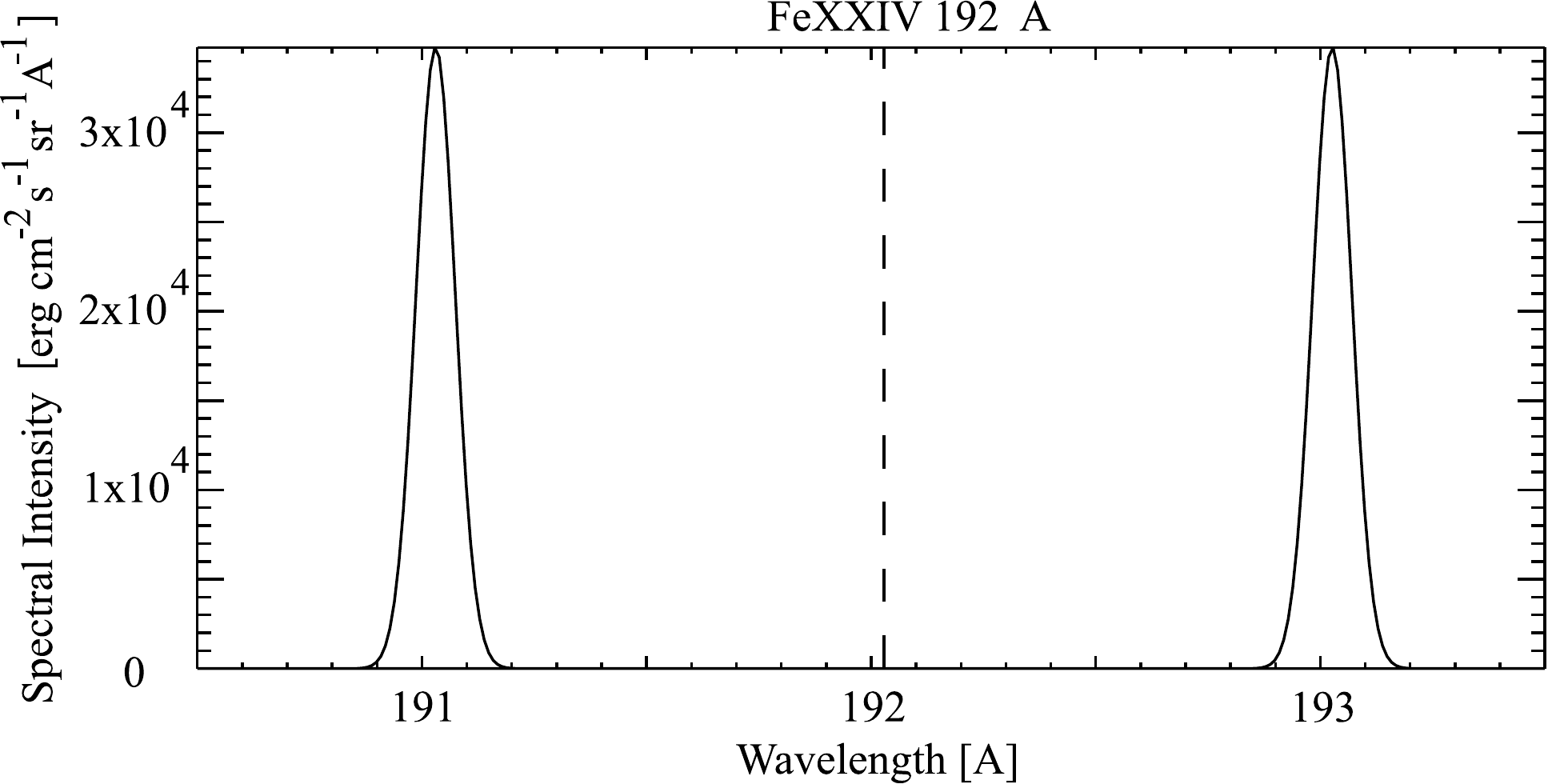}
\caption{ Spectral intensity of \ion{Fe}{24} along y=0 in Figure 6.}
\end{figure}

\begin{figure}
\epsscale{0.9}
\plotone{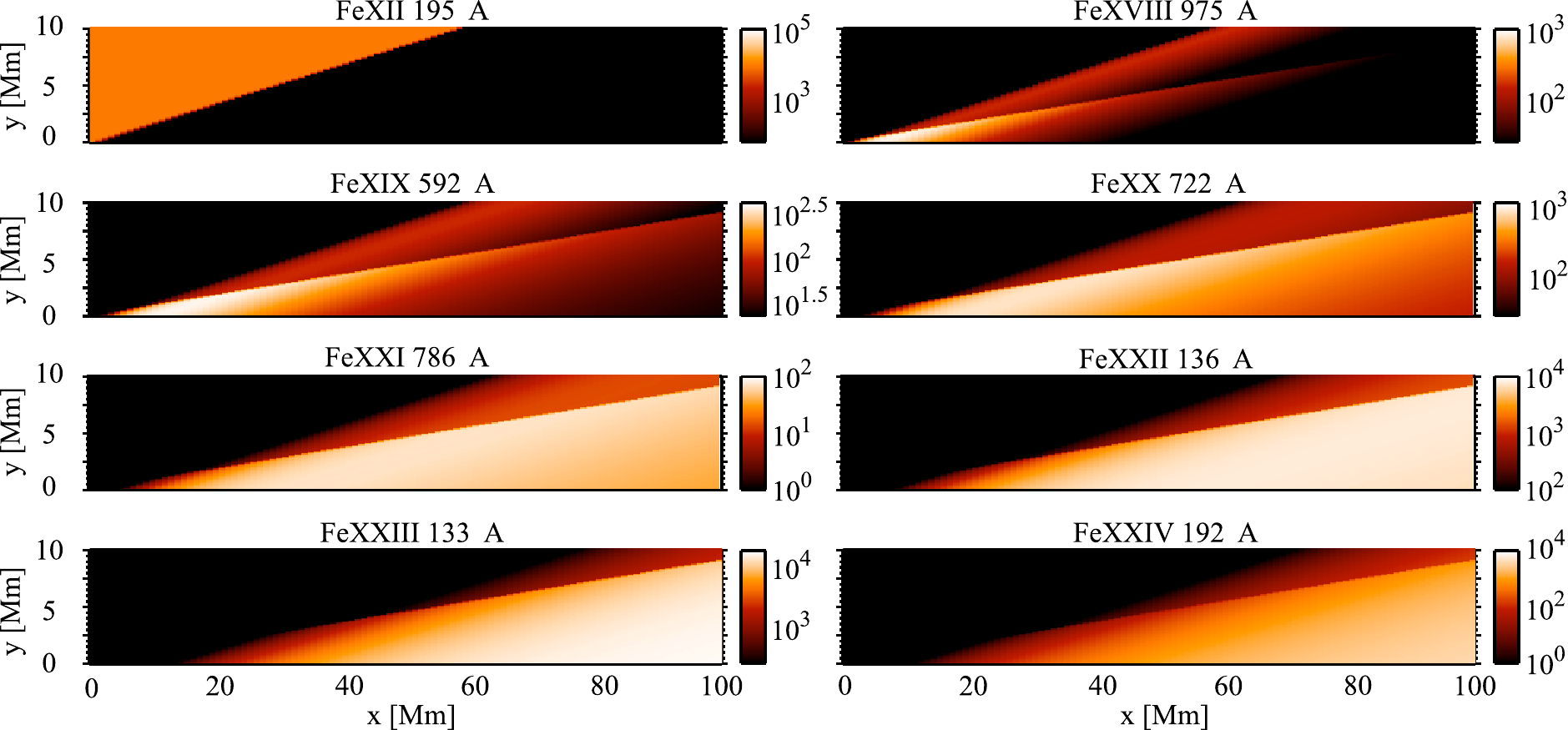}
\caption{ Result of time-dependent ionization in magnetic reconnection region with thermal conduction effect (Run4). }
\end{figure}

\begin{figure}
\epsscale{0.7}
\plotone{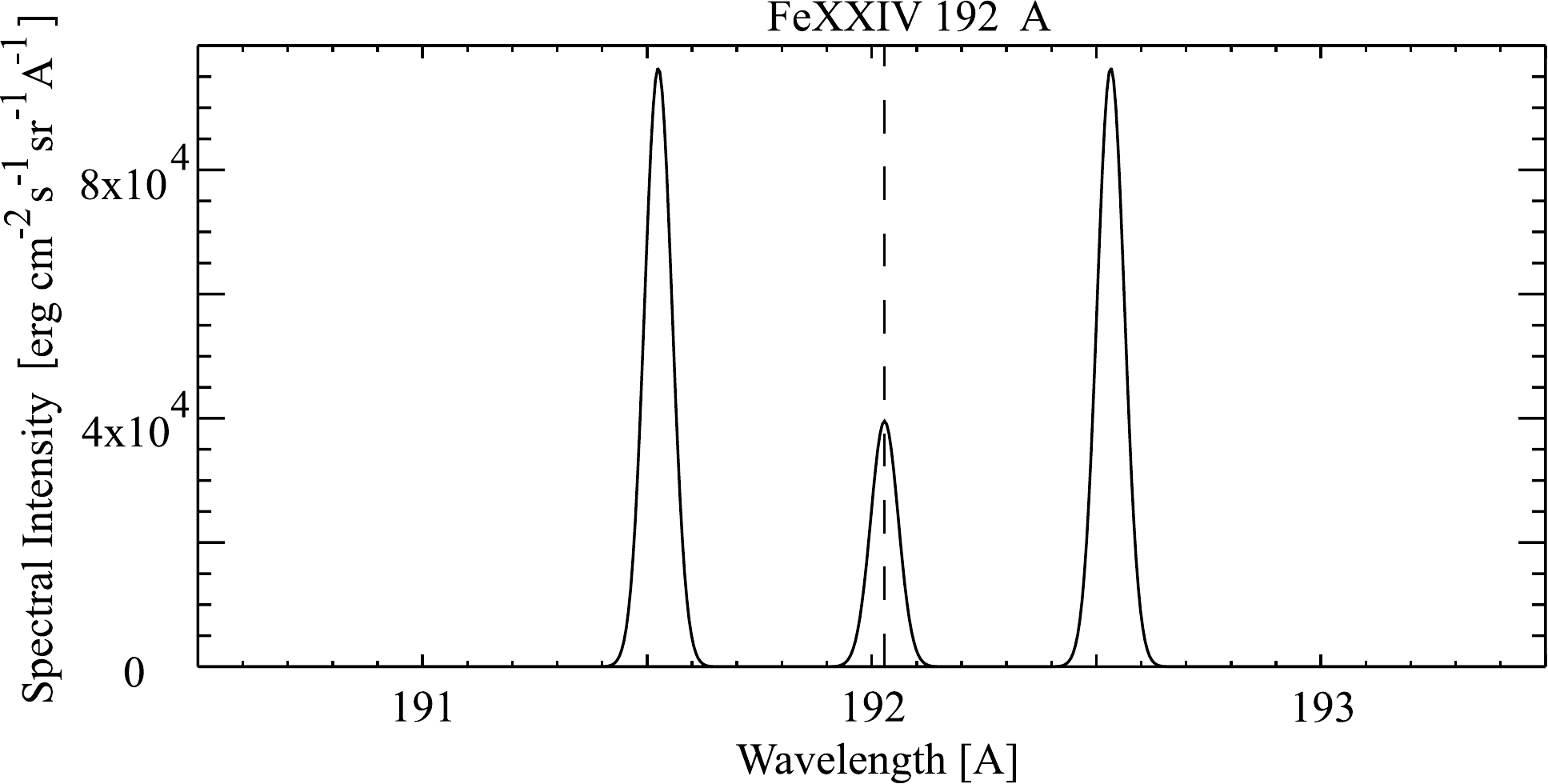}
\caption{ Spectral intensity of \ion{Fe}{24} along y=8.5 in Figure 8. }
\end{figure}

\begin{figure}
\epsscale{0.5}
\plotone{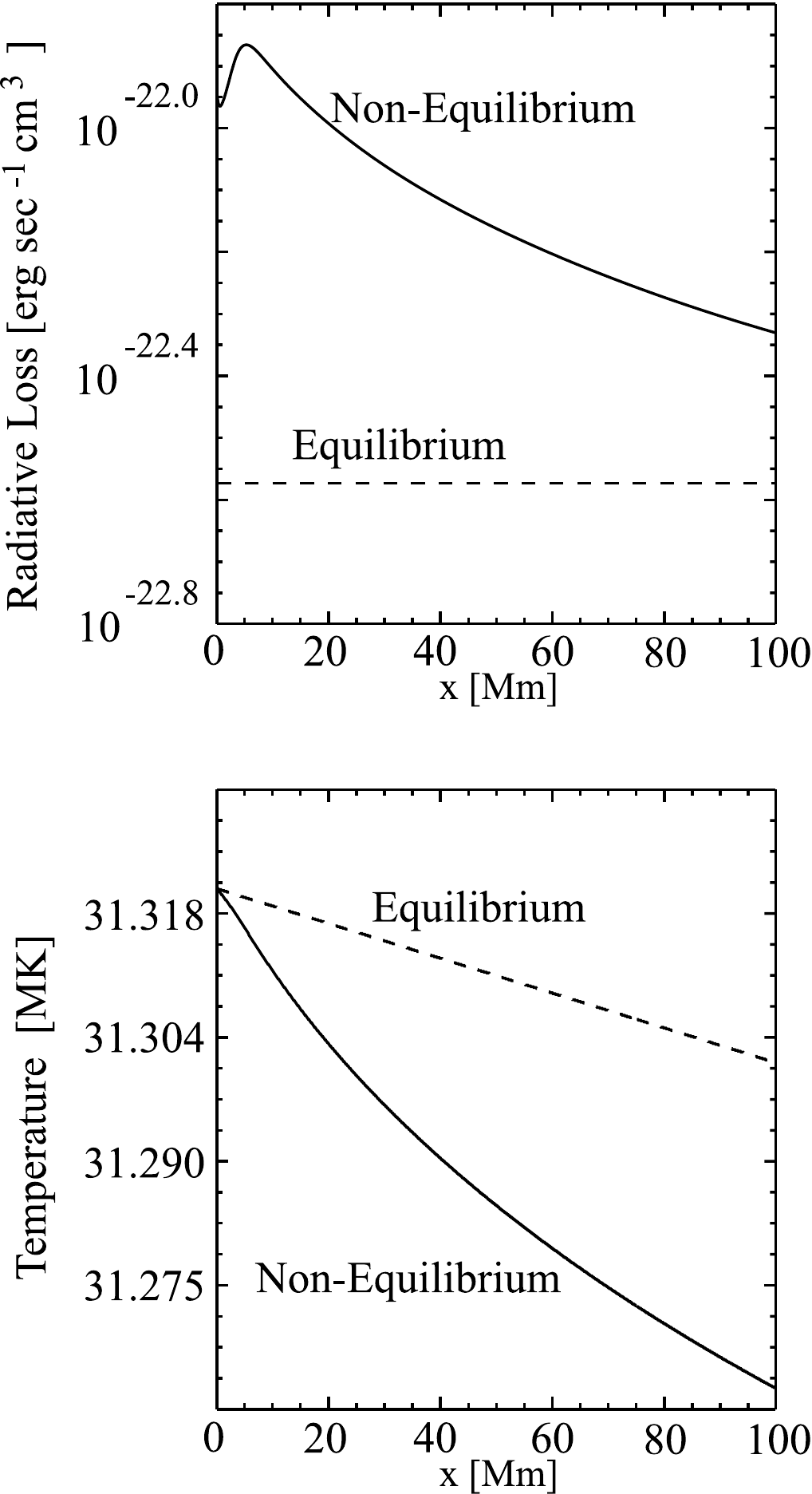}
\caption{ Spatial variation of the radiative energy loss rate and electron temperature in the downstream of the slow-mode shock. The solid lines show the results including the time-dependent ionization process, and the dashed lines show the results with ionization equilibrium assumption. Note that the vertical axis is logarithmic scale.}
\end{figure}

\begin{figure}
\epsscale{0.4}
\plotone{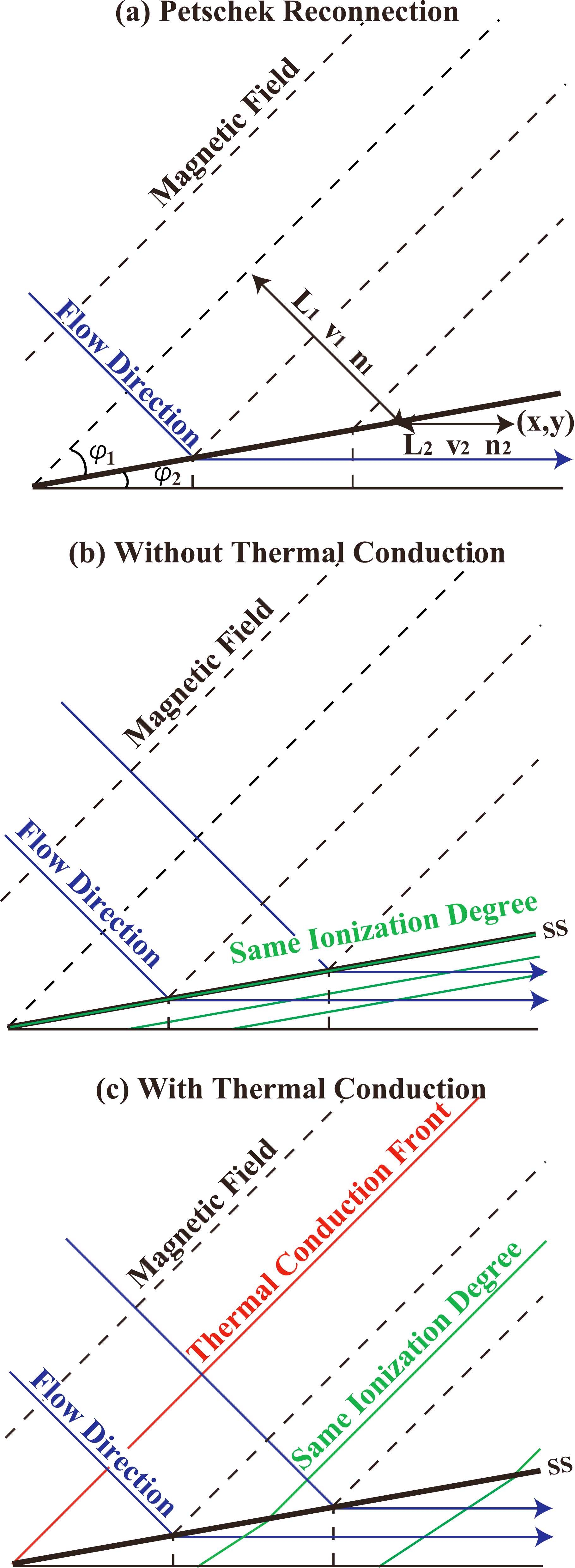}
\caption{ Schematic illustration of ionization in magnetic reconnection without and with thermal conduction.}
\end{figure}

\begin{figure}
\epsscale{0.5}
\plotone{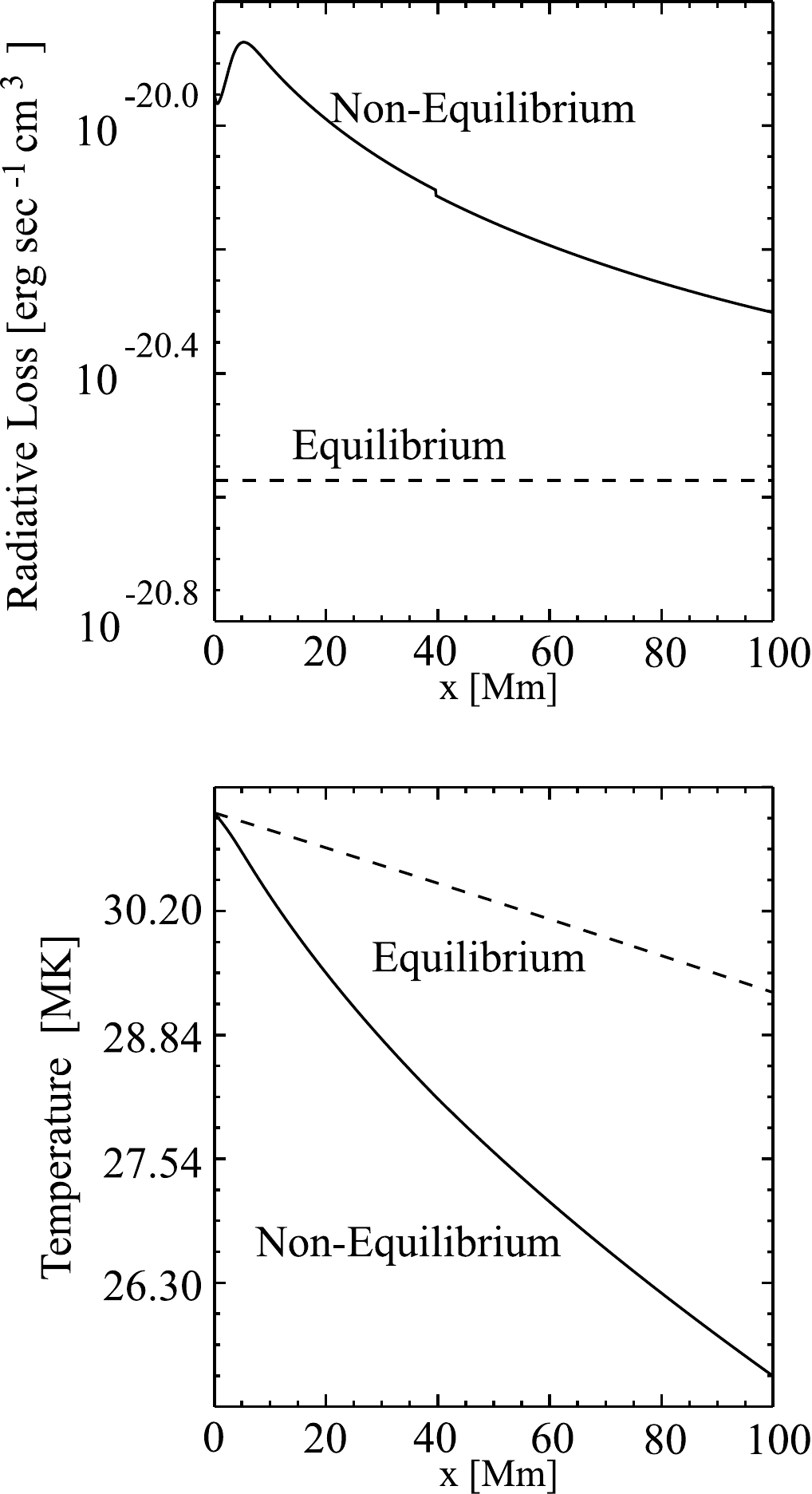}
\caption{ Spatial variation of the radiative energy loss rate and electron temperature in the downstream of the slow-mode shock with the assumption that the iron abundance is 100 times higher than the typical observed in the average corona. The figure format is the same as Figure 10.}
\end{figure}

\end{document}